# PERCEPTUAL CGAN FOR MRI SUPER-RESOLUTION

*Sahar Almahfouz Nasser\*, Saqib Shamsi\*\*, Valay Bundele\*, Bhavesh Garg\*, and Amit Sethi\**

\* Indian Institute of Technology Bombay, Mumbai, India.
\*\* Whirlpool, Pune, India

**ABSTRACT**

Capturing high-resolution magnetic resonance (MR) images is a time consuming process, which makes it unsuitable for medical emergencies and pediatric patients. Low-resolution MR imaging, by contrast, is faster than its high-resolution counterpart, but it compromises on fine details necessary for a more precise diagnosis. Super-resolution (SR), when applied to low-resolution MR images, can help increase their utility by synthetically generating high-resolution images with little additional time. In this paper, we present a SR technique for MR images that is based on generative adversarial networks (GANs), which have proven to be quite useful in generating sharp-looking details in SR. We introduce a conditional GAN with perceptual loss, which is conditioned upon the input low-resolution image, which improves the performance for isotropic and anisotropic MRI super-resolution.

*Index Terms*— Deep-learning, Super-resolution, MRI, GAN, Projection discriminator

## 1. INTRODUCTION

A magnetic resonance imaging (MRI) scan is very helpful for non-invasive medical diagnoses. However, an MRI is not an option for medical emergencies and is very difficult to do for pediatric patients. Moreover, MRI equipment cost goes up with its resolution. Furthermore, the scan time is also directly related to the resolution of an MRI, which makes it susceptible to patient motion. A 4x reduction in the degradation of spatial resolution would result in a 4x faster imaging time, but at the cost of the fine-grained details in the images that aid in making better diagnostic decisions. AI-based single image super-resolution (SISR) methods can reduce the time and cost of MRI acquisition by using low-resolution (LR) input images and quickly producing synthetic high-resolution (HR) images. The challenge is to produce the finer details that are close to those of the underlying HR images.

We propose a 3D SR method based on deep neural networks and generative adversarial networks (GANs) for MRI volumes that outperforms bicubic interpolation and state-of-the-art methods. We improve various individual components n the GAN framework to get an overall boost in the perceptual quality of super resolved MRIs. We study the impact of various architectural and loss function choices for both isotropic and anisotropic SR cases. Our contribution is threefold:
1. We constructed a 3D VGG-style network [1] for the MRI classification task, then we used this pretrained network, which has already learned to encode the perceptual and semantic information from the classification task, to compute the perceptual loss [2] between the real and and super resolved MRI volumes.
2. We proposed a conditional GAN that discriminates between the real and the super resolved images conditioned on the low-resolution MR images.
3. We did an extensive ablation study to determine the combination of the generator, the discriminator and the loss functions that works the best for MR super-resolution.

## 2. RELATED WORK

Super-resolution is an ill-posed inverse problem as it might have more than one solution for the high resolution image x given its low-resolution counterpart y [3]. The forward relation is described as follows:

$$y = (x * k)\downarrow_s + n \qquad (1)$$

where k is a blurring kernel, $*$ is convolution, $\downarrow s$ represents the down-sampling by factor s $\in$ R+, and n is the added noise.

An image super-resolution method can be based on interpolation, reconstruction, or learning. Although interpolation-based methods are simple and fast, they tend to produce blurry outputs. By contrast, reconstruction-based methods generate sharp-details and rich outputs by assuming some underlying prior knowledge to confine the solution space. Such algorithms become time-consuming as the scale factor increases. Learning-based methods outperform the previous ones in reconstruction detail, accuracy, and computation speed.

Within the set of learning-based methods, we further compare the proposed method with those based on deep learning only, because these have become the state-of-the-art. In 2015 Dong et al. [4] proposed super-resolution convolutional neural network (SRCNN). This was the first deep learning-based method for single image super-resolution which has led to a dramatic leap in this domain. SRCNN is a very shallow network; it adds details to the interpolated version of the LR image. Similarly, a polynomial neural network and using a CNN on the wavelet decomposition of images have also been proposed for image super resolution. [5, 6] Unlike SRCNN, fast super-resolution convolutional neural network (FSRCNN) [7] upsamples the feature maps at the end using a nearest-neighbor interpolation-based deconvolutional layer, which speeds the training and reduces the blur in the output image. Kim et al. [8] demonstrated that using a very deep super-resolution network (VDSR) further improves the quality of the output image. To boost the convergence, the authors used a high learning rate and gradient clipping.

However, the aforementioned methods suffer from the problem of vanishing gradient. To address this issue, many researchers have proposed architectures based on ResNet and DenseNet, such as super resolution ResNet (SRResNet) [9], enhanced deep residual network (EDSR) [8], and super resolution deep net (SRDenseNet) [10]. Zhang et al. [11] proposed residual channel attention network (RCAN), which exploits the existence of the short and long skip connections of the residual-in-residual (RIR) blocks [11] to train the very deep neural networks.

The increasingly impressive results of the GANs in generating images encouraged Wang et al. to explore their power in generating realistic textures for single image super-resolution [9]. That is, a generator is used to produce the super-resolved images, while a co-trained discriminator tries to discriminate between synthesized and real high-resolution images in order to improve the generator. Though the generated images had sharper details compared to the results of the existed super-resolution methods, this method produces unrealistic but sharp-looking artifacts. To counter the generation of unrealistic details, the use of different architectures, adversarial loss, and perceptual loss has been proposed [12].

### 3. PROPOSED METHOD

We propose a deep learning-based method with an end-to-end pipeline that learns the mapping of a low resolution 3D MRI to its high resolution counterpart.

For this study, we used two different generator architectures that have been introduced for 2D SISR and adapted them to 3D SR. We also introduced a 3D version of projection discriminator [13], and found that it benefited the training process as well as the overall results. To our knowledge, this is the first time that a 3D projection discriminator has been used for MRI super resolution.

The following subsections describe the model architectures of the generator and discriminator that we used.

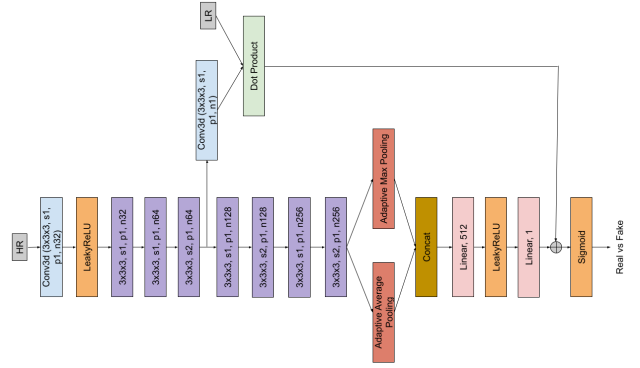

**Fig. 1:** Architecture of the projection discriminator: We compute the dot product between the LR MRI and a feature map of the HR MRI, which is then added to the output of the activation of the last fully connected layer [13].

### 3.1. Projection Discriminator

Conditional GANs (cGANs) [14] have shown immense usefulness for the task of conditional image generation. Unlike in standard GANs, the discriminator of a cGAN discriminates between the generator's conditional distribution and the target conditional distribution of generated samples x conditioned upon a paired input y. A projection based discriminator to incorporate conditional information into the discriminator has been proposed [13]. They showed the effectiveness of the model over other approaches of feeding conditional information to the discriminator, such as concatenating the conditional information with either the input or the feature map learned by one of the intermediate layers in the network. We adapted their framework for MRI by constructing a projection discriminator using 3D-convolutions.
For super-resolution, the following formulation was used for the discriminator function:

$$f(x, y; \theta) = \sum_{i,j,k,l}(y_{ijkl} F_{ijkl}(\Phi(x; \theta_{\Phi}))) + \psi(\phi(x; \theta_{\phi}); \theta_{\psi}) \quad (2)$$

where $x \in R^{R_{H1} \times R_{H2} \times R_{H3} \times R_{H4}}$ is the high resolution MRI and $y \in R^{R_{L1} \times R_{L2} \times R_{L3} \times R_{L4}}$ is the low resolution MRI and $F(\varphi(x; \theta\varphi)) = V * \varphi(x; \theta\varphi)$ for a convolutional kernel V and

convolutional operator ∗. The architecture of projection discriminator is shown in Fig. 1

## 3.2. Generator Architectures

We tested the following two generator architectures.

**SRResNet:** Our first generator architecture is a fully convolutional one inspired by [9]. This generator consists of a 3D-convolutional block followed by eight residual blocks, a point-wise convolution and a $3 \times 3 \times 3$ convolution operation to reduce the number of channels before upsampling so that the architecture becomes memory efficient, followed by an upsampling block and the network ends with another convolutional block. We also use a global skip connection. We use a combination of an upsample and a convolution operation instead of using a sub-pixel convolution followed by a reshape operation (as was used in [15]) to avoid checkerboard artifacts. There is no activation function applied at the end of the network, see figure 2a.

**Residual Dense Network:** Our second generator architecture is a residual dense network (RDN), inspired by [16]. A residual dense block (RDB) consists of a sequence of dense layers followed by a skip connection. The architecture of RDN is similar to SRResNet which just the residual blocks replaced by RDBs, see figure 2b.

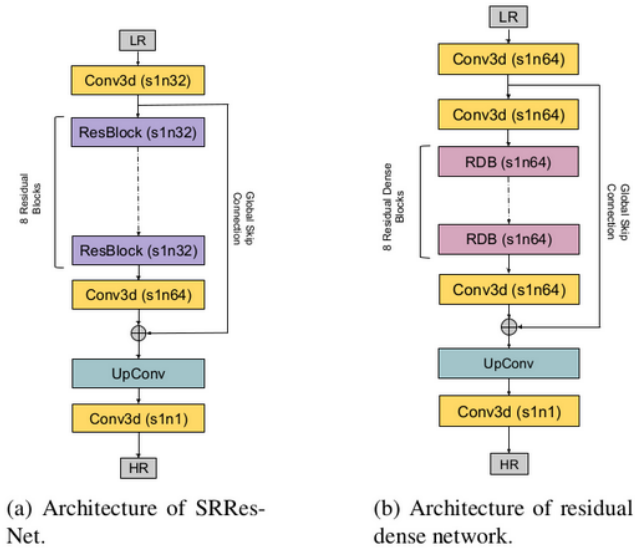

(a) Architecture of SRRes-Net.  (b) Architecture of residual dense network.

**Fig. 2**: Generator architectures.

## 4. DATA AND EXPERIMENTS

The Super-resolution of Multi-Dimensional Diffusion MRI (Super MUDI) dataset [17] contains the data of four healthy human subjects with ages range between 19 and 46 years. For each subject 1,344 MRI volumes are provided. The imaging device was clinical 3T Philips Achieva Scanner (Best, Netherlands) with a 32-channel adult head coil.

The Super MUDI Challenge comprises two tasks: isotropic, and anisotropic super-resolution. The names of these tasks were derived from the acquisition strategies of the low-resolution MRI data. The objective of using two down-sampling strategies is to compare the combinations of the down-sampling methods and the super-resolution approaches that can best to be used in a clinical scheme to obtain simulated high-quality and high-fidelity MRI images while reducing the acquisition time. In the anisotropic subsampling the volume has high in-plane resolution (2.5mm $\times$2.5mm), but thick axial slice (5mm), while in the isotropic subsampling the volume has low resolution (5mm) in all the directions.

For our experiments, we use one subject each for training and validation, and two for testing.

### 4.1. PERCEPTUAL LOSS

As the perceptual loss considers the structural and content similarities between the real and generated volumes at different scales, it improves the reconstruction of HR volume. To study its effect on the performance of our proposed method, we constructed a 3D VGG network trained on MR data. This network comprises five convolutional blocks and a classification head. The 3D kernels of all the convolutional layers are of sizes ($3\times3\times3$). The number of filters increases by a factor of two as we go deeper starting from 64 filters at the first layer. Every convolutional block consists of a convolutional layer, a 3D batch normalization layer, and a ReLU activation function. The classification head contains three fully connected layers of dimensions 512, 128, and 3.

The task of the VGG network is to classify the input image into one of three classes T1, flair, or diffusion MRI volume to learn the semantic information and later transfer this knowledge to the generator when training it to minimize the perceptual loss.

Our dataset contains 23 T1 volumes and the corresponding flair volumes of 23 subjects from the dataset [18], and 250 diffusion volumes randomly selected from Super MUDI dataset [17]. As our dataset was very imbalanced, we trained the network using class-balanced softmax cross-entropy loss [19], described as follows:

$$CB_{softmax}(z,y) = -(1-\beta/1-\beta^{n_y})log(exp(z_y)/\sum_{j=1}^{C}exp(z_j)) \quad (3)$$

where $z = [z_1, z_2, ..., z_C]^T$ is the prediction of the model, C is the total number of classes, ny is the total number of training samples corresponding to class y, $\beta \in [0,1)$ is a hyperparameter.

## 5. RESULTS

Our method (SRResNet+PD+PL) outperformed bicubic interpolation (winner of the Super MUDI challenge) as well as two state-of-the-art SR techniques to which we compared DCED[20] and 2D-ESRGAN [21] – for both isotropic and anisotropic SR of MRI images. Qualitative results are shown in Figures (3,4) and quantitative ones in Table 1.

We also performed an ablation study with various components introduced in Section 3. Projection discriminator (PD) alone was not sufficient to guarantee improved results. Adding the perceptual loss (PL) to train the combined architecture (SRResNet+PD) improved the results for all the evaluation metrics by an appreciable extent for both isotropic
and anisotropic super-resolution. However, perceptual loss (PL) deteriorates the performance when added to some other combinations of the generator and the discriminator but it preserves the semantic information. As Figure 4 depicts there are missing details without PL (red box for RDN+PD), but not with PL (RDN+PD+PL) even though the latter's metrics are lower. Finally, the combination of RDN and SD, which is equivalent to 3D ESRGAN, surpassed 2D ESRGAN. Moreover, the use of 2D architectures with interpolation creates unwanted artifacts in the slices, which is evident in the Figures 3 and 4 for 2D-ESRGAN.

**Table 1:** Results for isotropic and anisotropic super-resolution. SD, PD and PL refer to standard discriminator, projection discriminator and perceptual loss respectively.

| Experiment | Isotropic SSIM | Isotropic PSNR | Anisotropic SSIM | Anisotropic PSNR |
|---|---|---|---|---|
| Bicubic | 0.63 | 10.55 | 0.63 | 21.71 |
| SRResNet + SD | 0.68 | 21.08 | 0.74 | 25.45 |
| SRResNet + PD | 0.52 | 18.21 | 0.75 | 24.62 |
| SRResNet + SD + PL | 0.57 | 21.71 | 0.11 | 1.43 |
| **SRResNet + PD + PL** | **0.83** | **31.31** | **0.90** | 33.01 |
| RDN + SD | 0.66 | 23.60 | 0.82 | 28.70 |
| RDN + PD | 0.76 | 27.45 | 0.37 | 11.80 |
| RDN + SD + PL | 0.36 | 3.88 | 0.60 | 19.30 |
| RDN + PD + PL | 0.47 | 6.04 | 0.32 | 2.95 |
| 2D-ESRGAN + Bilinear | 0.51 | 21.49 | 0.49 | 21.94 |
| DCED | 0.77 | 29.39 | 0.82 | **33.01** |

## 6. CONCLUSION

We proposed a perceptual cGAN for 3D MRI super-resolution, which outperformed bicubic interpolation (winning algorithm of the Super MUDI challenge) for reconstructing HR images from downsampled LR images. We showed the benefit of adding the perceptual loss (PL) to projection discriminator and 3-D SRResNet generator for recovering the finer details and producing perceptually more plausible images. We found that PL does not always improve performance because the perceptual and adversarial losses preserve the semantic information (structure and content), while per-pixel losses such as MSE preserve the style information (color, texture, and common patterns). Thus, training the generator on a combination of MSE, adversarial loss, and PL drives the generator to focus on the semantic information at the cost of the style information. However, the projection discriminator penalizes the differences in style to preserve the stylistic features, and forms a good combination for the PL that focuses on semantic information. However, the selection of the generator plays an important role too, which should be explored further.

SR strategies that can reconstruct clinically-relevant details in HR images are very promising for reducing MRI acquisition time. The clinical validity of the details generated by SR methods for medical images remains to be further, as the popular quantitative metrics – PSNR and SSIM – are clinically agnostic.

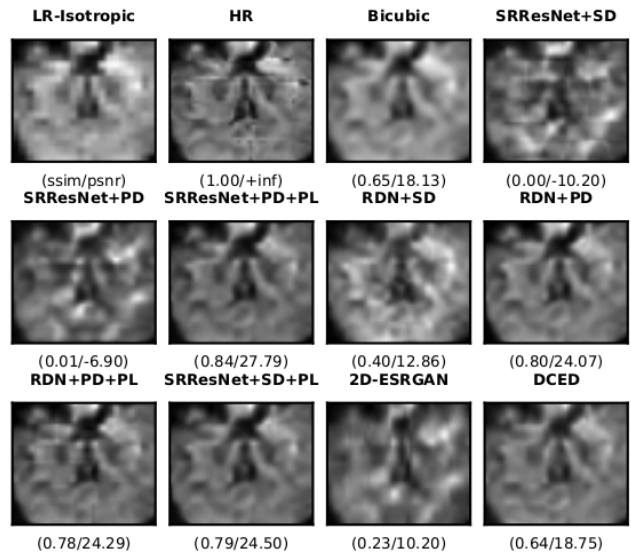

**Fig. 3:** Visualization of the results of the isotropic task for a slice of one of the patient data in the test set.

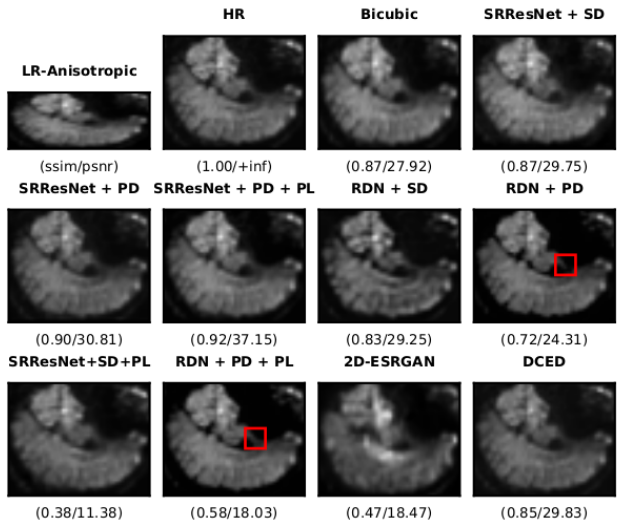

**Fig. 4**: Visual comparisons for the anisotropic task for a slice of one of the patient data in the test set.